\documentclass[a4paper, aps, prl, superscriptaddress, preprint]{revtex4-1}
\usepackage[utf8]{inputenc}
\usepackage[english]{babel}
\usepackage{graphicx}
\usepackage{ulem}
\usepackage{amsmath}
\usepackage{amsfonts}
\usepackage{tabularx}
\usepackage{float}
\restylefloat{table}
\begin{document}

\title{Vibrational surface EELS probes confined Fuchs-Kliewer modes}
\date{\today}

\author{Hugo Lourenço-Martins}
\affiliation{Laboratoire de Physique des Solides, CNRS, UMR8502, Bâtiment 510, Université de Paris-Sud, France}

\author{Mathieu Kociak}
\affiliation{Laboratoire de Physique des Solides, CNRS, UMR8502, Bâtiment 510, Université de Paris-Sud, France}

\begin{abstract}
Recently, two reports \cite{Krivanek2014,Lagos2017} have demonstrated  the amazing possibility to probe vibrational excitations from nanoparticles with a spatial resolution much smaller than the corresponding free-space phonon wavelength using electron energy loss spectroscopy (EELS). While Lagos et al. \cite{Lagos2017} evidenced a strong spatial and spectral modulation of the EELS signal over a nanoparticle, Krivanek et al. \cite{Krivanek2014} did not.  Here, we show that discrepancies among  different EELS experiments as well as their relation to optical  near- and far-field optical experiments \cite{Dai2014} can be understood by introducing the concept of confined bright and dark Fuchs-Kliewer modes, whose density of states is probed by EELS. Such a concise formalism is the vibrational counterpart of the broadly used  formalism for localized surface plasmons \cite{OUYANG1989,1997Garcia,abajo2008,Boudarham2012}; it makes it straightforward to predict or interpret phenomena already known for localized surface plasmons such as environment-related energy shifts or the possibility of 3D mapping of the related surface charge densities \cite{Collins2015}.
\end{abstract}

\maketitle

\begin{figure}[bthp]
    \includegraphics[width= \columnwidth]{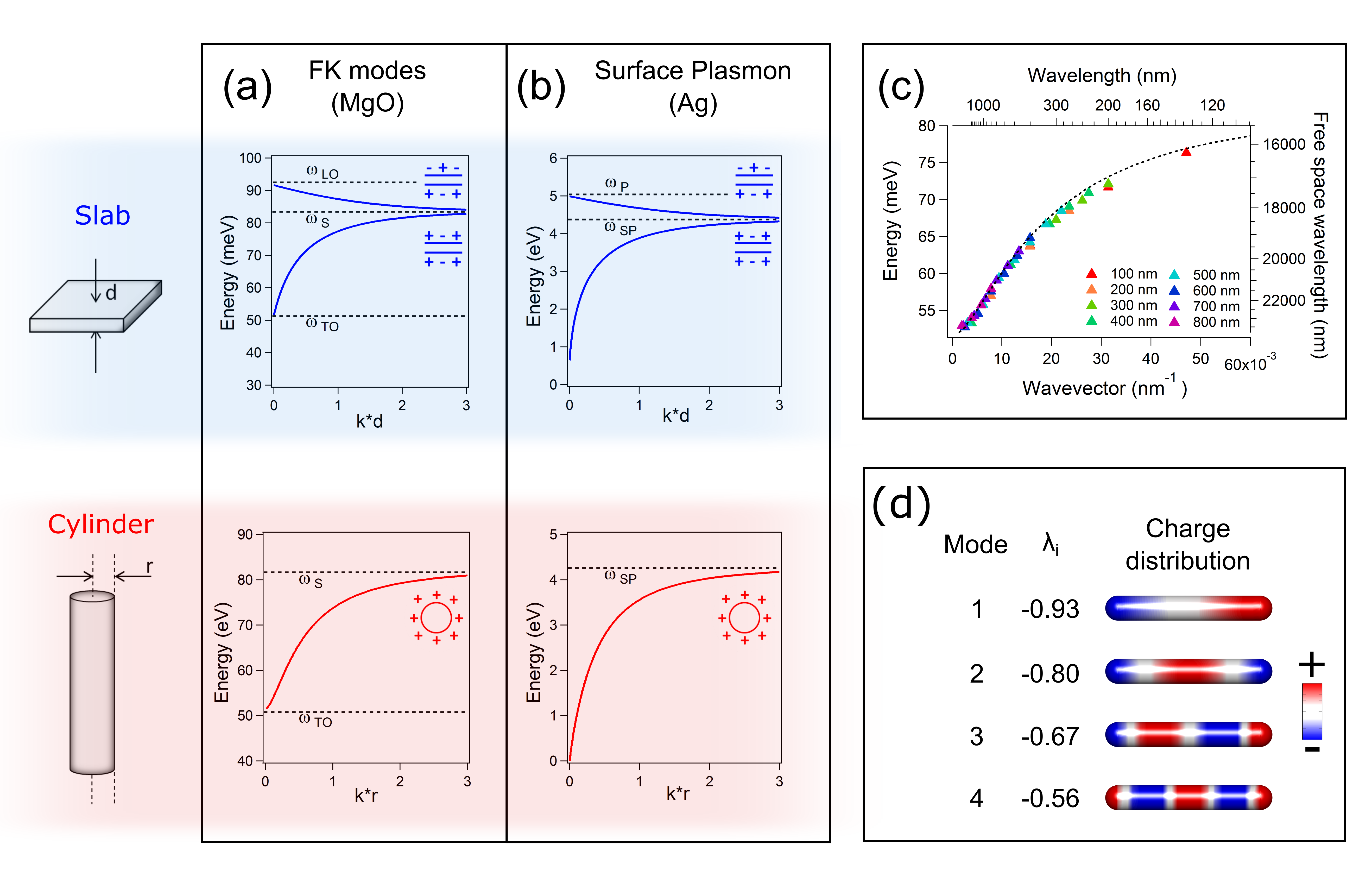}
    \caption{Analogy between Fuchs-Kliewer modes and surface plasmons modes. a:  Dispersion relation of the Fuchs-Kliewer modes for a slab of thickness $d$ (top) and a cylinder (bottom) of radius $r$ made up of MgO. The charge symmetry of the modes is sketched in inset. For the cylinder, only the rotationally invariant modes branch is shown, as the other modes are essentially not dispersing \cite{Kociak2014a}. b: Same for SP modes in silver. c: dispersion relation for the cFK in a series of  nanorods of different lengths (diameter is 30 nm). d: Surface eigencharges distribution for cFK of a nanorod, with the given mode orders and eigenvalues $\lambda_i$. }
    \label{fig:dispersion}
\end{figure}

Electron energy loss spectroscopy experiments consist in sending a free electron beam onto a sample of interest and retrieving information on its excitations through the analysis of the energy lost by the electron beam. It can essentially be performed without spatial resolution at low electron energy (HREELS) or with a sub-angström resolution in a scanning transmission electron microscope (STEM).
In a pioneering work, Ibach \cite{Ibach1970} used HREELS to analyse the vibrational excitations of a $ZnO$ surface. He could retrieve the measured value of the surface phonons energy $\omega_s$ within what was later called the local continuum dielectric model (LCDM) \cite{Lambin2003}. This simple and powerful model relies on the assumption that the local dielectric constant $\epsilon(\omega)=\epsilon(\omega,q=0)$ (where $\omega$ is the energy and $\epsilon(\omega)$ is equal to its value at zero transferred momentum $q$) is sufficient to describe electromagnetic excitations in a finite system. In Ibach's simple geometry, $\omega_s$ was such that $\epsilon(\omega_s)=-1$.  Fuchs and Kliewer demonstrated the amazing efficiency of the LCDM to describe more complicated geometries, such as slabs \cite{Fuchs1965}  and infinite cylinders \cite{FKrevue}. Already in these simple systems, the electromagnetic coupling between surfaces induces surface phonon splitting in so-called Fuchs-Kliewer (FK) with different charge distribution symmetries (Figure~\ref{fig:dispersion}a). Most materials dielectric constants can be described in the optical phononic range with a Drude-Lorentz model requiring the sole knowledge of the longitudinal and transverse optical phonon energies ($\omega_{LO}$ and $\omega_{TO}$) and the value of the dielectric constant at large energy values ($\epsilon_{\infty}$) of the bulk material, see Annex. One sees in Figure~\ref{fig:dispersion}a that FK modes disperse  as a function of the transferred wave-vector from $\omega_{TO}$ or $\omega_{LO}$ and converge to $\omega_s$ at large transferred wavevector.

The Fuchs-Kliewer work has been extended with an impressive success \cite{Lambin2003} to the description of surface plasmons (SP) in simple systems such as slabs and cylinders \cite{FKrevue,RUPPIN1990} (see Figure~\ref{fig:dispersion}b). Stimulated by the development of the research on plasmons in nanoparticles systems, several simulation schemes basically relying on the LCDM (boundary element model, BEM \cite{1997Garcia,2002Abajo,Hohenester2009} and discrete dipole approximation \cite{Geuquet2010}) have been extensively used to simulate optical and EELS spectra dominated by localized SPs \textit{confined} on nanoparticles. BEM simulations have been recently extended to the phonon range for STEM-EELS  \cite{Lagos2017} using the  MNPBEM \cite{Hohenester2014} implementation.

Now, beyond their unique simulation capabilities, LCDM derived theories have offered a deep understanding of localized SP physics. In particular, they made explicit the link between STEM-EELS and optical near-field spectroscopies as both are related to the electromagnetic local density of states (EMLDOS) \cite{abajo2008,Boudarham2012}, and showed that EELS is related to the extinction cross-section for dipolar modes \cite{Losquin2015,Losquin2015a}.

 The goal of this paper is to show how the reasoning once made to explain SP confinement in nanoparticles and interpret STEM-EELS experiments can now be used to rationalize the interpretation of surface STEM-EELS vibrational experiments in nano-objects and predict new physical effects.

In the following, we will introduce the confined FK (cFK) modes as surface phonons whose properties are mostly defined by the classical confinement that they experience in particles much smaller than the free-space equivalent wavelength. In this sense, if normal phonon modes are conceptually related to bulk plasmon modes and FK modes (also known as surface phonons) to surface plasmons,  cFKs are the phononic counterpart to localized SPs. For the sake of simplicity we will neglect retardation in the following. As we will show, this is justified by the relatively small sizes  of phononic nanoparticles studied in the literature \cite{Krivanek2014,Lagos2017}. A rigorous definition of the cFK modes can then be given in the quasi-static (QS) approximation using a modal decomposition form, first introduced in the case of confined SP \cite{OUYANG1989,1997Garcia,Boudarham2012}, see Annex. cFKs are then defined as a set of eigencharges $\sigma_i$ and eigenvalues $\lambda_i$, $i$ being the mode index. In the general case, $\lambda_i$, which depends only on the geometry of the nanoparticle, has to be determined numerically, and corresponding eigenenergies can be deduced through a simple implicit relation between $\lambda_i$ and the energy dependent dielectric constant (see Annex).  In the case of a model Drude-Lorentz dielectric constant, a general expression for the cFK eigenergies is (see Annex):
\begin{equation}\label{cFK}
	\omega_i=\sqrt{\frac{\epsilon_{\infty}\omega_{LO}^2(\lambda_i+1)-\omega_{TO}^2(\lambda_i-1)}{\epsilon_{\infty}(\lambda_i+1)-(\lambda_i-1)}}
\end{equation}

cFK energies lie between the bulk LO and TO energies, as $-1<\lambda_i<1$ \cite{OUYANG1989}, and we directly see that the energy of two well-known FK modes for an infinitely thin slab,  describing the charge-antisymmetric and -symmetric modes (see Figure~\ref{fig:dispersion}a), are retrieved for $\lambda_i=\pm 1$. In addition, other simple cases can be straightforwardly deduced.   $\lambda_i=0$ corresponds to the above-mentioned surface phonon \cite{Ibach1970}  case ($\epsilon=-1$) with eigenenergy $\omega_s=\sqrt{\frac{\epsilon_{\infty}\omega_{LO}^2+\omega_{TO}^2}{\epsilon_{\infty}+1}}$ in a Drude-Lorentz model, $\lambda_i=-1/3$ \cite{Kociak2014a} corresponds to the dipolar mode of a sphere ($\epsilon=-2$, $\omega_i=\sqrt{\frac{\epsilon_{\infty}\omega_{LO}^2+2\omega_{TO}^2}{\epsilon_{\infty}+2}}$). 

To exemplify the interest of this approach, we start with the case of nanorods that has been widely investigated in surface plasmon physics \cite{2011Novotny}, and especially by EELS \cite{2011Rossouw,Alber2011}. The simplicity of the structure makes it easy to understand the intimate link between shape and modes structures, and we adapt it here to the case of a phononic material following arguments for localized SPs found in \cite{Kociak2014a}. Modes in a nanorod of radius $r$ and length $L$ are similar to the FK modes of the infinite rod, except that the confinement restricts the available wavevectors to multiple of $1/2L$. This is exemplified in Figure~\ref{fig:dispersion}c where the discrete modes dispersion relation, simulated for a large set of nanorods lengths, overlaps the one of an infinite rod. Such modes are the cFK modes of the nanorod. The cFK modes disperse between  $\omega_{TO}$ and $\omega_s$, in analogy with the corresponding dispersion in localized SP in nanorods restricted between $0$ and $\omega_{sp}$ \cite{Kociak2014a}. Similarly to the corresponding localized SP modes, each mode with eigenvalue $\lambda_i$ corresponds to an oscillation of the surface eigencharge, as depicted in Figure~\ref{fig:dispersion}d. We see on the prototypical case of a nanorod that the QS approximation is much more justified for cFK than for localized plasmons for objects of same sizes:  the length (top scale in Figure~\ref{fig:dispersion}d) of a typical nanorod is much smaller than the equivalent free-space wavelength of the cFK (right scale in Figure~\ref{fig:dispersion}d). Another difference is the pile up of low order modes for long nano-antennas close to $\omega_{TO}$ which is obviously absent for localized surface plasmons.

\begin{figure}[bthp]
    \includegraphics[width= \columnwidth]{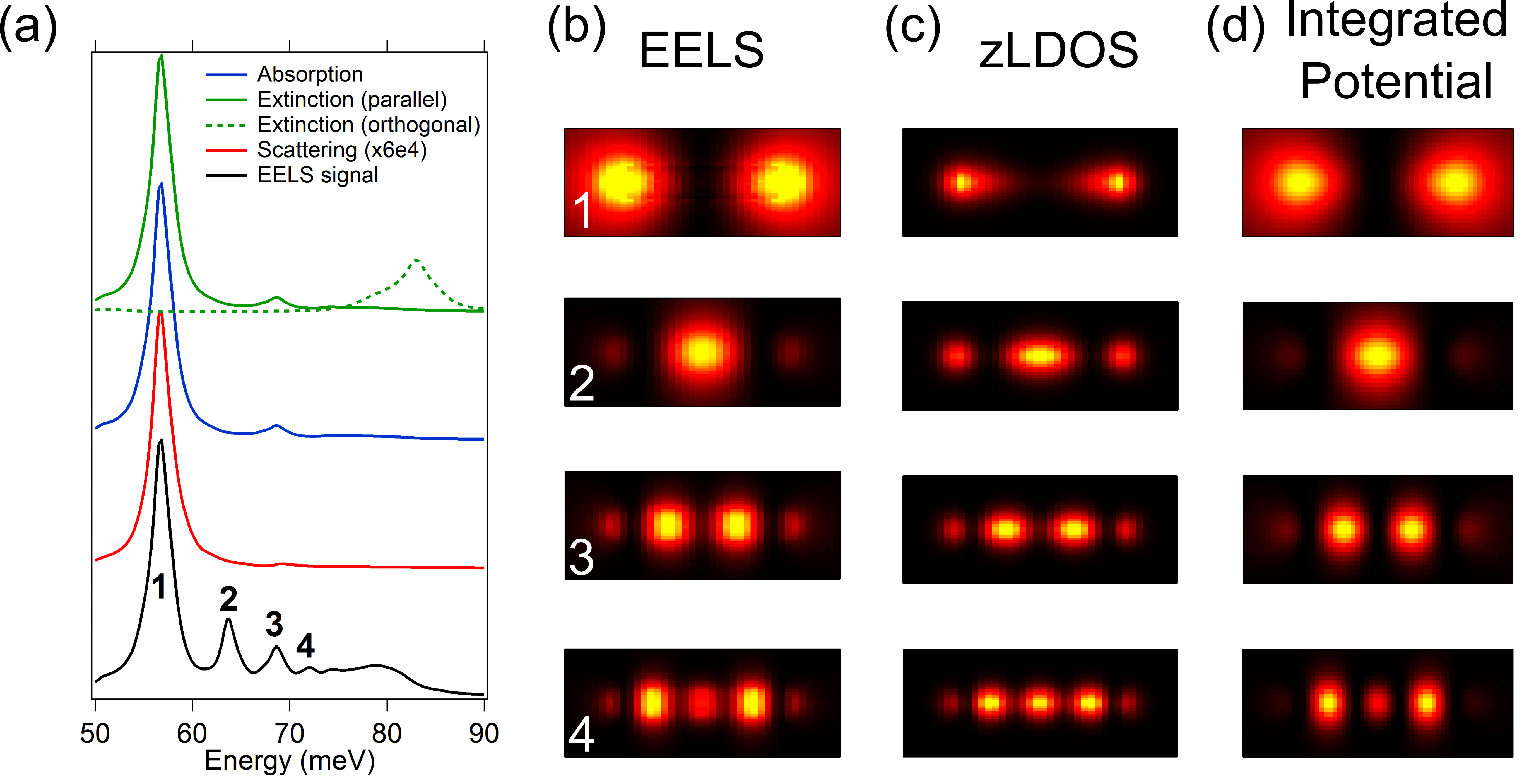}
    \caption{Optical cross-sections, EELS, EMLDOS and eigenpotentials for the cFK in a nanorod of MgO.  a. Simulated optical cross sections for an incoming beam propagating perpendicular to the nanorod axis, and  EELS spectrum for an electron beam  located 10nm away from one tip of the nanorod. All spectra have been shifted for clarity. Optical cross-sections scales are the same for extinction and absorption, and multiplied by $6.10^4$ for scattering. The polarization of the electrical field is parallel to the nanorod axis, except for the dotted line curve. The nanorod is 200 nm in length and 30 nm in diameter b. EELS maps for the four first modes of the nanorod. c. Corresponding zEMLDOS maps taken at z=10 nm from the surface of the rod. d. Corresponding z-integrated eigenpotentials.}
    \label{fig:eelsrod}
\end{figure}

Figure~\ref{fig:eelsrod}a presents one EELS spectrum simulated for beam impinging 10 nm away from one tip of a MgO rod of 200 nm long and 30 nm in diameter. The simulations performed in the full retarded approximation and using an experimental dielectric constant as an input \cite{Hofmeister2003} reveal a series of peaks. As seen on Table~\ref{tab:antennas}, a direct comparison of their energy values with that of the cFK deduced from Eq. \ref{cFK} based on the sole knowledge of the $\lambda_i$, $\omega_{TO}$, $\omega_{LO}$ and $\epsilon_{\infty}$ shows an almost perfect agreement. This validates conceptually our approach, and also allows to use a simple EELS modal decomposition (see Annex \ref{eq:EELS}) for EELS simulations. 

In Figure~\ref{fig:eelsrod}a, we also compare EELS to macroscopic optical quantities such as the absorption, extinction and scattering cross-sections calculated in the retarded approximation. As in the case of EELS, the spectra do not peak at the normal modes energies $\omega_{LO}$ and $\omega_{TO}$. Instead, they are dominated by the cFK modes, in analogy with the well-known case of a slab spectrum dominated by the FK modes \cite{Fuchs1965} or more generally for an ensemble of nanoparticle \cite{FUCHS1975}. This is particularly justified from the modal decomposition of the cross-sections, see Eq. \ref{eq:ext} in Annex and \cite{Losquin2015}: the optical cross-sections are proportional to a spectral function peaking at the dipolar cFK modes energy.  Contrary to the case of EELS, only the dipolar modes are observable (but a very slight contribution from the third order mode). The spectra obviously show a large dependence on the incoming polarization. For polarizations along the nanorod axis, the dipolar mode of the low energy branch is excited. For a polarization perpendicular to it, the dipolar modes of the other branches, almost all arising at $\omega_s$ \cite{Gomez-Medina2008}, are excited, see Figure~\ref{fig:eelsrod}a. This points to the fact that EELS is sensible to both bright (i. e. optically active) and dark (i.e. not optically active) cFKs, in contrast to optical far-field techniques.

\begin{figure}[bthp]
     \includegraphics[width= \columnwidth]{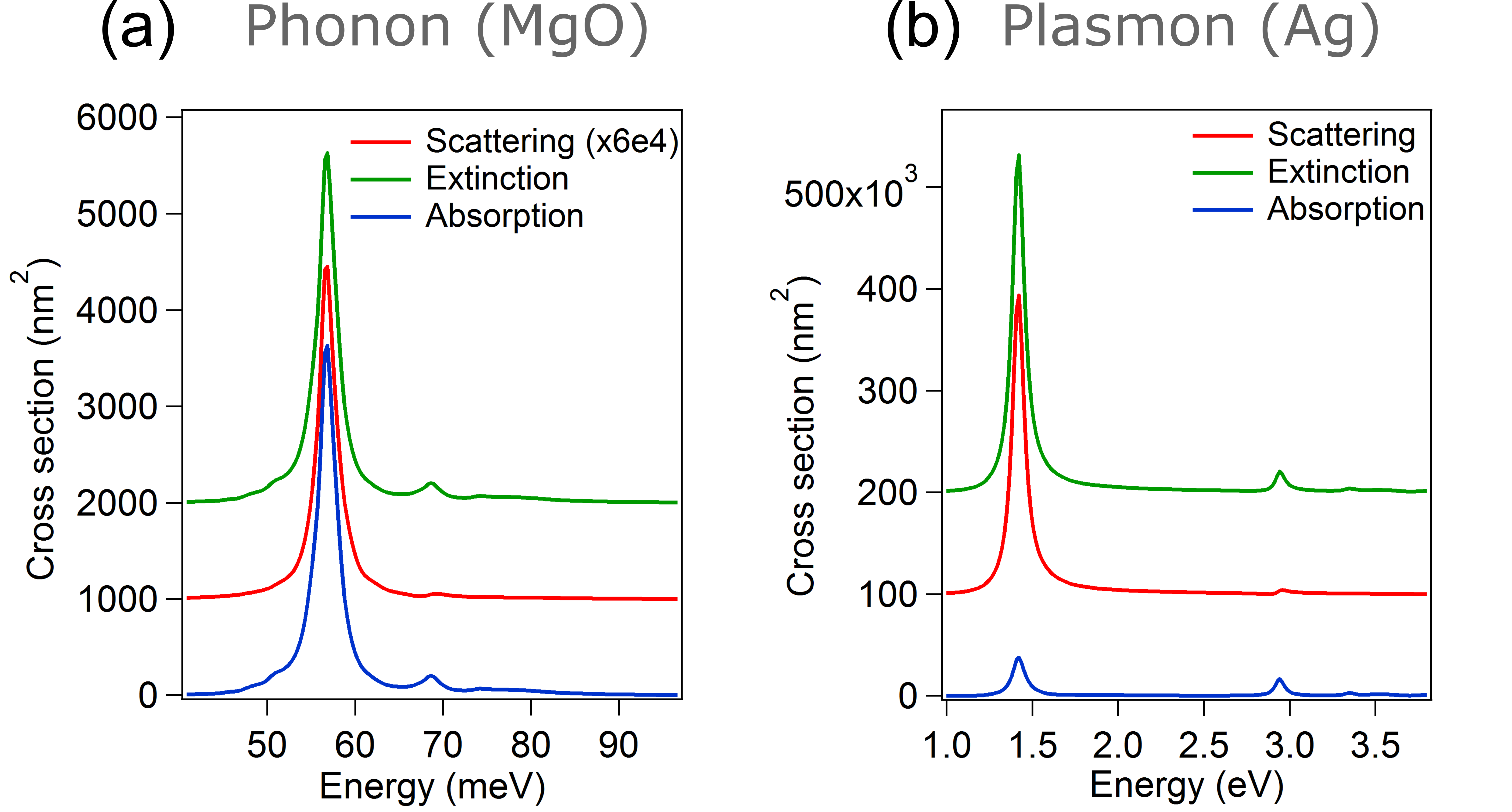}
     \caption{Optical extinction, absorption and scattering cross-sections for a. A MgO nanoantenna and b. A silver nanoantenna. Both antennas have the same size (200X30 nm). Note the relative cross-section values.}
     \label{fig:antennas}
 \end{figure}

 Obtaining truly dark (non-emitting/absorbing) localized SPs is difficult due to the relatively large sizes of plasmonic particles \cite{Losquin2015} with respect to the corresponding free space wavelengths. In contrast, for the cFKs where the QS approximation is justified for much larger particles sizes, almost only dipolar modes are  bright. We note that the scattering cross-section is several order of magnitude smaller than the extinction one. This is basically related to the fact that, other things being equal, the ratio between scattering and extinction scales as $1/\omega^3$, where $\omega$ is the energy of interest. This makes extinction and absorption cross-sections almost identical at the low energy of the phonon regime, making EELS very close to the absorption cross-section for dipolar cFK modes (see also the analytical proof in the Annex). We note that this contrasts with the case of a silver plasmonic nanorod of the same size (see Figure~\ref{fig:antennas}). In this case, scattering has a major contribution in the extinction cross-section.

We can now clarify the type of selection rules when exciting cFK optically or with electrons. To start with, in the QS approximation, only dipolar modes can be excited by a plane wave, and the electrical polarization of the plane wave must be aligned with the dipole direction. Away from the QS, similar symmetry arguments arise: even modes (mode 2 and 4 on Figure~\ref{fig:dispersion}d) cannot be excited by a plane wave with electrical field in the plane containing the axis of the nanoantenna, while odd modes (1 and 3) can be excited. Tilting the beam direction with respect to the antenna axis will break the symmetry and make it possible to also detect even order modes. More generally, for optical experiments, the selection rules are completely determined by the general symmetry of the surface charge distribution with respect to the plane wave direction and polarization.

 The interplay between the symmetries of the incoming electron electrical field  and the surface eigencharges  is different. As with optics, cFK modes are also probed by EELS, but  contrary to optics, EELS is sensible to all modes even in the QS approximation. Also, the symmetry of the surface eigencharges impacts rather the spatial distribution of the EELS signal. Indeed,  EELS maps (Figure~\ref{fig:eelsrod}b) closely resemble the EMLDOS projected along the electron propagation direction $z$ (zEMLDOS, Figure~\ref{fig:eelsrod}c), with the EMLDOS spatial and spectral distribution being essentially determined by the size, shape and symmetries of the object of interest. The resemblance between EELS and zEMLDOS is expected by analogy with the localized SP case, where also a general analytical relation between these two quantities can be determined  \cite{abajo2008}. Much as in the case of localized SPs \cite{Boudarham2012}, EELS  as well as near-field optical techniques do not map directly the eigencharges \cite{Guzzinati2017}. Rather, they map the related zEMLDOS, itself related to the z-projection of the electric eigenfield in the QS limit \cite{Hohenester2009,Boudarham2012}. An even more precise description of EELS of cFK in terms of electromagnetic quantities is given by the almost identity between EELS and the z-integrated eigenpotentials \cite{Horl2013}, see Figure~\ref{fig:eelsrod}d.

We can sum up the results exemplified on the nanorods but valid for any kind of phononic nanoobject.

First, \textit{surface} EELS and optical IR absorption, extinction and scattering are probing the same physical excitations, namely cFK. The symmetry of the cFK surface eigencharges, which depends on the global shape and symmetry of the subtending particle determines the coupling strength of the cFK with the probing electrons or photons. This is in stark contrast with IR absorption  or \textit{bulk} EELS  \cite{Dwyer2016,Forbes2016,Lagos2017}, which are probing \textit{normal} modes, which depend on local (atomic) symmetries, i. e. the bulk material properties.   This is also a main difference between our work, which relates surface vibrational EELS to the concept of \textit{EMLDOS}, and recent theoretical works describing the link between bulk EELS to the concept of phononic density of states (\textit{pDOS}). Again, pDOS is dependent on the atomic structure symmetry while EMLDOS is dependent on the  global (shape) symmetry of the nanoparticle. Also, for similar reasons, surface EELS is completely different to Raman spectroscopy which probes bulk properties of atomic oscillations, although following selection rules different to that of bulk IR absorption. Note that the LCDM can also be used to predict the bulk EELS experiments results through a term proportional to $-Im(1/\epsilon(\omega))$, giving essentially a peak at $\omega_{LO}$ in the Drude-Lorentz model. The intensity of the related peak maybe influenced by the screening at the surface, a phenomenon handled in the LCDM theory and known as "begrenzung"  effect \cite{Lagos2017}. There are however several limits explaining the need to develop dedicated theories for bulk phonons beyong the LCDM \cite{Dwyer2016,Forbes2016,Lagos2017}, related to  the interpretation of angular resolved experiments \cite{Dwyer2016,Forbes2016,Lagos2017}.   

Second, EELS maps are close to that obtained with the near-field optical measurement which are related to the EMLDOS \cite{Hillenbrand2002}, and map quantities close to the cFK electric eigenfields, and more precisely the  eigenpotentials, along the electron direction integrated on the electron beam path (see an analytical proof in Eq. \ref{eq:EELS} and \cite{Horl2013}). The typical spatial extent of the EELS signal is related to that of the EMLDOS, and almost identical to that of the integrated eigenpotentials.

 Third, due to the large free space wavelength of the cFK compared to typical dimensions of nano-objects, the QS approximation holds essentially true for sub-micron nanoparticles, and any nanoparticle can be described by a series of eigencharges and related $\lambda_i$ that only depends on the shape of the nanoparticle. 
 
 In addition, this theory works well for understanding cFK, but will obviously fail to describe long-wavelength, propagating surface phonons that may arise in the particular case of very large particle or slabs. In the case of slabs or infinite cylinders, however, alternatives rigorous retarded theories exist \cite{Fuchs1965}. The differences in the predictions between a quasi-static (such as presented here) and retarded formalism weakly affect lowest energy, charge-symmetric modes that are usually dominant in slabs and cylinders. 
 
 Also, a rigorous modal decomposition of all relevant EELS and optical quantities for arbitrary shaped nanoparticles (see e.g Eq. \ref{eq:EELS} and \ref{eq:ext}) is possible simplifying both the understanding and predictions of surface EELS experiments. Finally, the formalism presented here is not specific to the Drude-Lorentz model (except of course Eq. \ref{cFK}  and \ref{eq:g}). Therefore, any situation where a local dielectric constant can be deduced, either theoretically or experimentally, can be handled. For example, \textit{ab- initio} models of the IR dielectric constant of a crystal of molecules could be computed, and re-injected in our model for interpreting quantitatively the experiments, just as recently performed by Radtke et al. \cite{Radtke2017} in the case of a planar interface to interpret results on guanine crystals \cite{Rez2016}. With all these considerations in mind, we are in the position to synthesize observations made in the literature on surface phonons in terms of FK modes or cFK modes.


 Krivanek et al. \cite{Krivanek2014} reported the first observation of vibrational signatures with STEM-EELS. Among other, they reported a resonance at 173 meV on a $\approx$ 50 nm thick sheet of hexagonal boron nitride (hBN), and a resonance at 138 meV in an $\approx$ 30 nm thick $SiO_2$ slab. The resonances energy did not change as a function of the electron beam position  whether it was impinging the objects or in vacuum close to them. The 173 meV resonance was attributed to the LO normal mode of hBN, and the other compared to  IR results without further assignment. Following the reasoning of this paper, one can rationalize these results, see also Table \ref{tab:slab}. The 173 meV (hBN) modes and 138 meV ($SiO_2$) are likely to be  charge-symmetric (lower branch in Figure~\ref{fig:dispersion}, $\lambda_i$ close to $-1$) FK modes. Indeed, with the help of equation \ref{cFK} (see Table \ref{tab:slab}), one can directly deduce that their energies are  between the $\omega_{TO}$ and $\omega_{s}$  (and very close to $\omega_{TO}=169.5$ meV in the case of hBN) but largely different from $\omega_{LO}$, see Table \ref{tab:slab}. For symmetry reasons, the dipole strength of the charge-antisymmetric mode vanishes with the thickness of the slab \cite{2002Stephan}. It might explain why this mode was not reported in \cite{Krivanek2014}. On the other hand, as summarized in Table \ref{tab:slab}, Batson and Lagos \cite{Batson2017} reported the measurement of two peaks on an h-BN flake, the first at 187 meV (below $\omega_s$) and the second at 203 meV (above $\omega_s$). These are likely to be charge symmetric and antisymmetric modes reciprocally - as confirmed by preliminary simulations in \cite{Batson2017}-  for a slightly thicker slab (as the symmetric mode energy is at higher energy and the symmetric mode is still weaker but now  measurable).  It is worth noting that in these cases, the energy of the modes depends on the geometry and symmetry of the nano-object, and we expect of course the observation of thickness dependent modes when more experimental works will be available in the literature. Finally, no modes energy spatial variation has been reported on these two sorts of slabs \cite{Krivanek2014,Batson2017}. Recently, Schmidt et al. \cite{Schmidt2014a} showed that the plasmonic modes in thin objects with edges can be decomposed in slabs modes and edges modes independently. The slabs modes follow the infinite slabs dispersion relations, and edges the nanoantennas ones \cite{Campos2017}. The modes of lowest energy branches have the same charge symmetry with respect to the slab or cylinder mid-plane, so that the slab and edge lowest energy modes share the same symmetry. Translated to surface phonons in $SiO_2$ slabs it means that we should expect two different modes of same symmetry with respect to the slab mid-plane; however, both dispersion curves are very close (see e.g Figure~\ref{fig:dispersion}a), and for very thin objects both slabs and edge modes energy tend to a unique and same value ($\omega_{TO}$), making it difficult to detect experimentally any spectral or spatial variation except an intensity decrease in vacuum.

\begin{figure}[bthp]
    \includegraphics[width= \columnwidth]{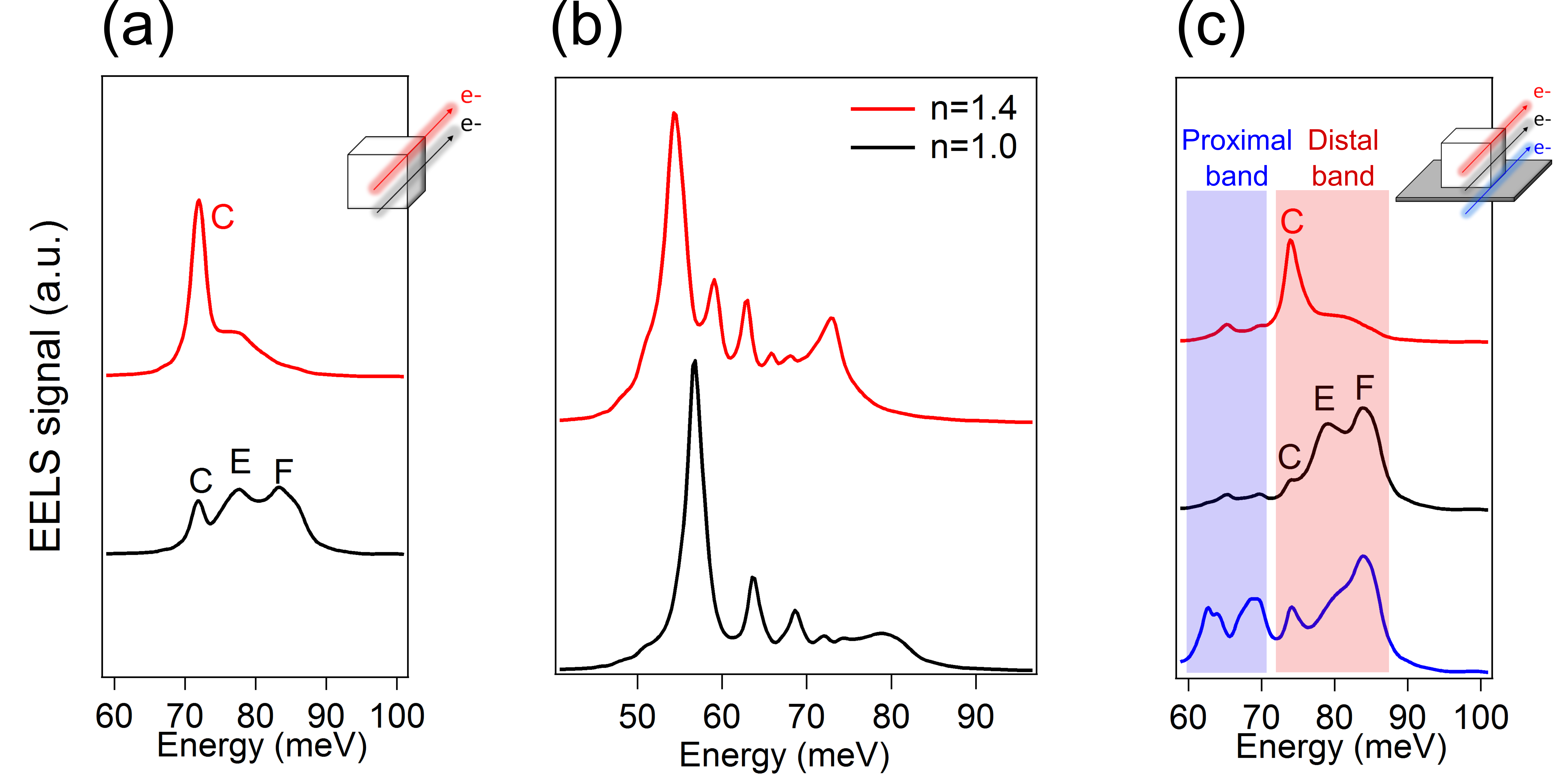}
    \caption{Dielectric environment effect. a. Simulated EELS spectra for a cube of MgO (100 nm edge long) in vacuum, exhibiting a corner (C), an edge (E) and a face (F) mode depending on the beam position. b. Simulated EELS spectra for a nanorod (200x30 nm) in vacuum (black) and embedded into a dielectric of refractive index equals to $1.4$. The beam is positioned at 10 nm from the tip of the nanorod in both cases.  c. Same simulations as in a., but for a cube deposited on a substrate of refractive index n=2.3. The former C, E and F mode split into two bands. The distal band is essentially consisting in a series of C, E, F modes   arising at almost the energy of the corresponding vacuum  modes, while the proximal band is shifted towards the $\omega_{TO}$ energy. Spectra corresponding to a given trajectory are indicated by their colors. }
    \label{fig:cubes}
\end{figure}

\begin{figure}[bthp]
    \includegraphics[width= 10 cm]{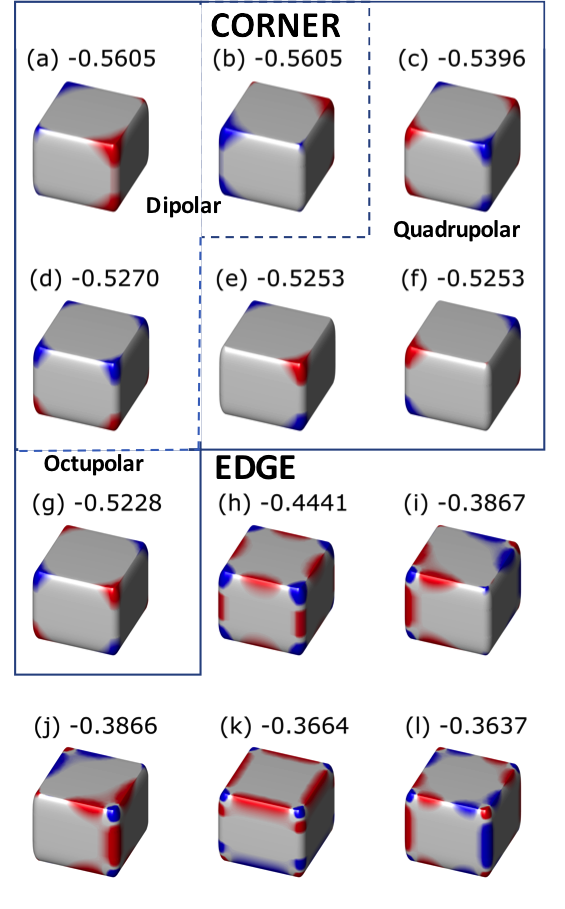}
    \caption{Modes symmetry for a cube in the quasi-static approximation. Values of $\lambda_i$ are given on top of the corresponding eigencharge distributions (red is minimum and blue maximum). a-g: corner modes. h-l: edge modes. Corner modes have been separated with respect to their symmetries.}
    \label{fig:cubesSI}
\end{figure}

  At the opposite, Lagos et al. \cite{Lagos2017} observed outside of MgO nanocubes an EELS signal with different energies and  clear spatial modulations. They identified essentially three modes (see also Figure~\ref{fig:cubes}a), a corner (C) one at lower energy, an edge (E) one and a  face one (F) at higher energies. All the modes could be simulated without taking into account any substrate. Table \ref{tab:cube} sum up Lagos' experimental and simulation results, as well as our simulations and the energies as deduced from Eq. \ref{cFK}. Our simulations are in good agreement with Lagos simulations and experimental results, not a strong surprise as our simulations and Lagos ones are performed with the same tool (MNPBEM), similar cube parametrization and the same full retarded approximation. More interestingly, we see in Table \ref{tab:cube} how well Eq. \ref{cFK} reproduces our simulations and Lagos's ones, themselves pointed to be in very good agreement with experiments (\cite{Lagos2017}).  Our theory gives however a stronger insight into the nature of the probed modes. In Lagos et al. \cite{Lagos2017}, modes are denominated through their EELS spatial distribution, with no discussion on their symmetries, which are known to be complex for cubes plasmons \cite{Zhang2011,Mazzucco2012}. Indeed, as shown in Figure~\ref{fig:cubesSI}, the corner mode can be decomposed in  dipolar, quadrupolar and octupolar contributions (see also Table \ref{tab:cube}) that are degenerated in the quasistatic approximation. Because one of its components is dipolar, the corner mode is likely to be bright  (ie theoretically measurable through an IR extinction experiment) although weakly scattering compared to a plasmonic cube of the same size.  Quite interestingly, the edge mode is in fact composed of a large number of cFKs of close $\lambda_i$, see  Table \ref{tab:cube} and SI. The symmetry of all these constituting modes makes the edge mode a dark one. Concerning the face mode, the number of polygons required for convergence was to high to deduce a definite value or set of values for $\lambda_i$. However, this highest energy mode has an energy very close to $\omega_s$ for MgO, corresponding to $\lambda_i=0$ (see Annex). This is expected from localized SPs analogy, as high momenta modes converge systematically to this value.

We now turn to a point which has not been considered so far but may have important implications for the interpretation of the forthcoming experiments. Indeed, the effect of the substrate, known to be essential in plasmon physics, has not been discussed in the context of surface vibrational STEM-EELS experiments. It is well-known that localized SP energy and spatial distribution drastically depend on the close presence of other materials, like a substrate or an embedding matrix. In Figure~\ref{fig:cubes}b, we show the effect of embedding a phononic nanorod into a material of constant dielectric constant different to one. It produces an expected redshift of the excitation, yet still constrained between $\omega_{TO}$ and $\omega_{LO}$. The case of a nanoparticle on a substrate  is more subtle. In particular, in the case of a nanocube, it is well-known from localized SP physics that the modes will split into  modes at low energy localized close to the substrate (proximal modes) and  at higher energy close to the vacuum (distal modes) \cite{Zhang2011}. In \cite{Lagos2017}, only the distal modes were reported, although both types of modes are actually predicted (see Figure~\ref{fig:cubes}). We note that the distal modes energies are very close to the mode of a free space cube, explaining the good agreement between our theory, Lagos and our simulations without substrate, and experimental results. Observation of the proximal band would however require a spectral resolution even better than actually available.
 
	Finally, the theory presented here can be extended to understand more complicated situations. This is in analogy with the success of the theory presented for localized SPs \cite{OUYANG1989,1997Garcia,abajo2008,Boudarham2012,Losquin2015a}, which has been extended to the 3D mapping of the EMLDOS \cite{Hoerl2015} or of the surface eigencharges \cite{Collins2015}, the simulation of the cathodoluminescence signals \cite{Losquin2015,Losquin2015a}, the interaction of surface excitations with phase-shaped incoming beams \cite{Guzzinati2017}, or the coupling between localized SP. Also, this model can be refined by developing a retarded model or a non-local approximation extension \cite{2010Garcia}.

\clearpage

\section{Annex}

\subsection{Modal form of the cFK, modal EELS and application to a Drude-Lorentz model}

Following \cite{OUYANG1989,1997Garcia}, the electromagnetic properties in the quasi-static approximation of an object of dielectric constant $\epsilon(\omega)$ in vacuum can be entirely determined by the set  $\{\sigma_i(\vec{s}),\lambda_i\}$, respectively the surface eigencharge and the eigenvalue for the mode $i$, $i$ being an integer and $\vec{s}$ the surface position vector. Actual eigenenergies can be determined through the dispersion relation $\lambda_i=\frac{1+\epsilon(\omega_i)}{1-\epsilon(\omega_i)}$. From this set, that can be determined numerically \cite{OUYANG1989,1997Garcia,Hohenester2014}, one can deduce all eigen-quantities such as the eigenpotential or the electrical eigenfield $\vec{E}_i(\vec{r})$ at all points $\vec{r}$, or any observable such as the EMLDOS $\rho_{\alpha\alpha}(\vec{r},\omega)$ (here $\alpha$ represents the projection direction):
\begin{equation}
	\rho_{\alpha\alpha}(\vec{r},\omega)=\frac{1}{2\pi^2 \omega}\sum_i Im(-g_i(\omega))|E_{\alpha}^i(\vec{r})|^2
\end{equation}

 the EELS probability (simplified here to the case where the beam is outside of the object of interest) \cite{Boudarham2012}:
\begin{equation}\label{eq:EELS}
	\Gamma(\vec{R}_{\perp},\omega)=\frac{1}{\pi \omega^2}\sum_i Im(-g_i(\omega))|E_z^i(\vec{R_{\perp}},\omega/v)|^2
\end{equation}
where $v$ is the speed of the electron, $z$ the direction of electron propagation, $\vec{R}_{\perp}$ the position of the beam in the plane perpendicular to $z$ and the extinction cross-section, which is equal to the absorption cross section in the QS limit reads \cite{Losquin2015}:
\begin{equation}\label{eq:ext}
	C_{ext}(\omega) \propto \sum_{i,d} A_{i} \omega Im(-g_i(\omega))
\end{equation}
where $A_i$ is a mode dependent prefactor, and the sum runs over the dipolar $d$ cFK modes only.

 $g_i(\omega)$ is a spectral function for mode i depending only on $\epsilon$ and $\lambda_i$ \cite{Boudarham2012} with imaginary part peaking at the cFK energy $\omega_i$.
  
  The above formulation clearly points out the fact that the EELS spectra are a superposition of cFK spectral functions weighted spatially by the modulations of the associated electrical eigenfields, and the close resemblance between EELS and EMLDOS, and the spectral similarities between EELS and extinction cross-section.
 In the case where the phonon response can be characterized  with  LO and TO energies $\omega_{LO}$, $\omega_{TO}$, a dissipation parameter $\Gamma$ and a dielectric constant at large energy $\epsilon_{\infty}$, a Drude-Lorentz form of the dielectric constant reads:

 \begin{equation}
 	\epsilon(\omega)=\epsilon_{\infty}(1+\frac{\omega_{LO}^2-\omega_{TO}^2}{\omega_{TO}^2-\omega^2+i\omega \Gamma})
 \end{equation}

then 
\begin{equation}\label{eq:g}
	Im(-g_i(\omega))=\frac{\Gamma \omega }{(\omega^2-\omega_i^2)^2+\Gamma^2\omega^2} 	\left[\frac{2(\omega_i^2-\omega_{TO}^2)^2}{\epsilon_{\infty}(\omega_{LO}^2-\omega_{TO}^2)(1+\lambda_i)}\right]
\end{equation}

the spectral function then takes the simple form of a lorentzian peaking at the cFK mode energy $\omega_i$ (solution of equation \ref{cFK}, this is the energy of the ith cFK in absence of dissipation), weighted by some energy independent prefactor. 

EMLDOS, EELS and absorption cross-section can be straightforwardly deduced from this expression of the spectral function.

The above deductions can be extended analytically to the case where the object of interest is embedded in a medium. Similar developments (see SI of \cite{Losquin2015a} or \cite{Hoerl2015}) can be done in the retarded regime assuming a model dielectric function.

\subsection{Analogy between localized SP and cFK modes}

From the point of view of the local continuum dielectric model, there is no functional difference between SPs and surface phonons, SP in slabs and cylinders and FK modes, and localized SP and cFKs, as long as the details of the dielectric constant are not disclosed. In the case where the SPs are described by a Drude model and the cFKs by a Drude-Lorentz model, the analogy between SPs and cFKs  can be simply made by replacing $\omega_{TO}$ by $0$, $\omega_{LO}$ by $\omega_p$ and $\epsilon_{\infty}$ by 1. Then, all calculations expressions presented in this paper can be compared to that for SPs, especially those found in \cite{Boudarham2012}. For example, one retrieves the familiar values of $\omega_p/\sqrt{2}$ and $\omega_p/\sqrt{3}$ for the surface and dipolar surface plasmons.

\subsection{Comparison of cFK energy values as computed through equation 1 and through retarded simulations}
\begin{table}[H]
\begin{tabular}{|l|l|l|l|l|l|}
\hline
 & Mode 1 & Mode 2  & Mode 3  &  Mode 4  & Surface\\
& $\lambda_i=-0.93$ & $\lambda_i=-0.8$   & $\lambda_i=-0.67$  &  $\lambda_i=-0.56$  & $\lambda_i=0$ \\ 
 & $\omega_1$ (meV) & $\omega_2$ (meV) & $\omega_3$ (meV)  &  $\omega_4$ (meV) & $\omega_s$ (meV)\\
 \hline
	from Eq.1 & 56.0& 63.4  & 68.7 & 72.3 & 83.1 \\
	Simulations (Figure \ref{fig:dispersion}d) & 56.8 & 63.6  & 68.6  & 72.0 & 82.9\\
	\hline
\end{tabular}
\caption{Comparison between energies values for the nanoantenna in Figure \ref{fig:dispersion}d  calculated with Equation 1 and as measured on the simulated spectra in  Figure \ref{fig:eelsrod}. Inputs for equation 1 are $\omega_{TO}=50.7 meV$, $\omega_{LO}=91.3$ meV, $\epsilon_{\infty}=3.01$ \cite{Hofmeister2003}. Simulations have been performed in the full retarded approximation, with the experimental dielectric constant found in \cite{Hofmeister2003}.}\label{tab:antennas}
\end{table}

\subsection{Normal mode, surface phonon and dipolar surface phonons for some materials} 
\label{sub:normal_mode_surface_phonon_and_dipolar_surface_phonon_for_hbn_sio__2_and_sic}
For $SiO_{2}$ and MgO, the energy of simple FK and cFK modes can be straightforwardly deduced from equation \ref{cFK} and the values given in the tables below. Limit analytical cases for the energy of the surface phonon ($\omega_s$), the charge-symmetric and -antisymmetric FK modes for an infinitely thin slab (converging to $\omega_{TO}$ and $\omega_{LO}$) and the dipolar mode for a sphere ($\omega_d$) have been given in the main text. Main values calculated with equation \ref{cFK} are given in the table below. 

The case of $hBN$ is a bit more involved, as hBN is an uniaxial anisotropic material. Nevertheless, the FK theory can be extended to anisotropic materials for slabs \cite{FKrevue}. The charge symmetric mode converge to the in-plane TO mode energy $\omega_{TO_{\perp}}$ and the charge anti-symmetric to the out-of-plane LO mode energy $\omega_{LO_{||}}$. The terminology $\perp,~||$ is related to the anisotropy axis. Likewise, the surface phonon energy will be a combination of in-plane and out-of-plane phonons energy given by the condition $\sqrt{\epsilon_{\perp}\epsilon_{||}}=-1$, with $\epsilon_{\perp}$ and $\epsilon_{||}$ the in and out of plane dielectric constant \cite{2001Kociaka}. We note that an HREELS study \cite{Oshima1997} reported a value for the LO mode of a single hBN sheet around $173$ meV, much as the value reported by \cite{Krivanek2014}. Given the similarities pointed out in the paper between HREELS and STEM-EELS and the symmetry arguments, the reported LO mode is most likely to rather be a charge-symmetric FK mode.

\begin{table}
\begin{tabular}{|l|l|l|l|l|l|l|}
\hline
	Material & $\epsilon_{\infty}$& $\lambda_i=-1$  & $\lambda_i=1$  &  $\lambda_i=0$  & $\lambda_i=-1/3$& slab \\ 
		 & &  $\omega_{TO}$ (meV) &  $\omega_{LO}$ (meV) &   $\omega_{s}$ (meV) &  $\omega_{d}$ (meV)& experimental (meV) \\ \hline
	$SiO_{2}$& 2.99 \cite{Fujii1992} & 134 \cite{Fujii1992} & 153 \cite{Fujii1992} & 143.8 & 140.6&138\\
	hBN  (in-plane)& 4.95 \cite{Geick1966} & 169 \cite{Geick1966} & 200 \cite{Geick1966} & 195 & &\\
	hBN (out-of-plane)& 4.1 \cite{Geick1966} & 187 \cite{Geick1966} & 197 \cite{Geick1966} & 195 & &\\
	hBN slab (\cite{Krivanek2014})& &  &  &  & &173\\
	hBN slab (experiment \cite{Batson2017})& &  &  &  & &187/203 \\
	hBN slab (theory \cite{Batson2017})& &  &  &  & &181/197 \\
	\hline
\end{tabular}
	\caption{Comparison of theoretical and experimental values for $\lambda_i=-1,1,0,-1/3$ (charge symmetric/antisymmetric FK modes for infinitely thin slabs or a cylinders), surface FK mode, dipolar spherical mode) and experimental values from \cite{Krivanek2014} and \cite{Batson2017}. In the later case, two modes (interpreted as charge symmetric, charge antisymmetric FK modes) are given.}\label{tab:slab}
\end{table}


\begin{table}[H]
\begin{tabular}{|l|l|l|l|l|l|l|l|}
   \hline
    Mode & \multicolumn{7}{c|}{\textbf{Corner}} \\
  \hline
   Symmetry & \multicolumn{3}{c|}{Dipolar} &\multicolumn{3}{c|}{Quadrupolar}&Octupolar\\
  \hline
   $\lambda_i$&-0.56 &-0.56 &-0.53 & -0.53 & -0.53 & -0.54 & -0.52 \\
   \hline
   $\omega$ (from Eq. 1)&72.3 &72.3 &73.1 & 73.1 & 73.1 & 72.8 & 73.1 \\
   
   \hline
   $\omega$ (simulations, this paper)&\multicolumn{7}{c|}{72.0} \\
   \hline
   $\omega$ (simulations, Ref. [2])&\multicolumn{7}{c|}{72} \\
   \hline
   $\omega$ (experiments, Ref. [2])&\multicolumn{7}{c|}{69} \\
   \hline\hline
    \hline
     Mode & \multicolumn{4}{c|}{\textbf{Edge}}&\multicolumn{3}{c|}{\textbf{Face}} \\
   \hline
    $\lambda_i$&-0.44 &-39 &... & All summed & \multicolumn{3}{c|}{N/A} \\
    \hline
    $\omega$ (from Eq. 1) &75.4 &76.5 &... & N/A& \multicolumn{3}{c|}{N/A} \\
   
    \hline
    $\omega$ (simulations, this paper)&\multicolumn{4}{c|}{77.7}& \multicolumn{3}{c|}{83.3} \\
    \hline
    $\omega$ (simulations, Ref. [2])&\multicolumn{4}{c|}{76}& \multicolumn{3}{c|}{83} \\
    \hline
    $\omega$ (experiments, Ref. [2])&\multicolumn{4}{c|}{72}& \multicolumn{3}{c|}{78} \\
	\hline
   
\end{tabular}
\caption{Comparison between energies values for the MgO nanocube modes calculated with Equation 1, from retarded simulations with experimental dielectric constant found in \cite{Hofmeister2003}, from retarded simulation in \cite{Lagos2017} and experimental results from \cite{Lagos2017}. Inputs for equation 1 are $\omega_{TO}=50.7 $ meV, $\omega_{LO}=91.3$ meV, $\epsilon_{\infty}=3.01$ \cite{Hofmeister2003}. Energies are given in meV units. Note the apparent discrepancy for the face mode values between simulations and experiments, proven in \cite{Lagos2017} to be an effect of finite spectral resolution in the experiments.}\label{tab:cube}
\end{table}


\subsection{Simulations} 
\label{sub:simulations}
Dispersion relations in Figure~\ref{fig:dispersion}a and b have been calculated using formulas from \cite{Rivacoba2000} and using a Drude model adapted to silver and a Drude-Lorentz adapted to MgO. All the other simulations have been carried out using the \texttt{MNPBEM toolbox}  \cite{Hohenester2014} using experimental values for the dielectric function of the MgO \cite{Hofmeister2003}. Figure~\ref{fig:dispersion}d has been calculated using the quasistatic eigensolver while Figure~\ref{fig:dispersion}c, Figure~\ref{fig:eelsrod}, Figure~\ref{fig:cubes} and Figure~\ref{fig:cubesSI} employ a retarded formulation of the Maxwell equations. Rods have been simulated using approximately 1000 polygons, cubes in vacuum with 5000 polygons and cubes on substrate with 5000 polygons as well. We simulated a $100$ nm length cube with approximately $6000$ polygons and calculated the corresponding eigencharges and geometrical eigenvalues $\lambda_i$ using the \texttt{plasmonmode} solver. The radii of curvature of the cube corners in the $xy$ plane are fixed at $3$ nm. The rounding in the $yz$ (resp. $xz$) direction is not precisely controlled within the \texttt{MNPBEM} toolbox \cite{Hohenester2014} (when using the \texttt{tripolygon} and \texttt{edgeprofile} functions). However we estimated the radius of curvature in these planes to be much shorter than 3 nm. Because of the slight asymmetry of the mesh, the three dipoles (resp. quadrupole and edge dipolar) are not slightly degenerated, see $\lambda_i$ values on figure \ref{fig:cubesSI}.

\section{Acknowledgements}

This work has received support from the National Agency for Research under the program of future investment TEMPOS-CHROMATEM with the reference ANR-10-EQPX-50.


\bibliographystyle{apsrev4-1}

\end{document}